\documentclass{pasj00}
\draft

\begin{document}
\SetRunningHead{T. Mizuno et al.}{Two Ultraluminous X-Ray Sources in NGC~1313}
\Received{2006/08/04}
\Accepted{2001/01/01}

\newcommand{\blue}{\textcolor{blue}}
\newcommand{\red}{\textcolor{red}}
\newcommand{\green}{\textcolor{green}}

\title{Suzaku Observation of Two Ultraluminous X-Ray Sources
in NGC~1313}



\author{
T. \textsc{Mizuno}, \altaffilmark{1}
R. \textsc{Miyawaki}, \altaffilmark{2}
K. \textsc{Ebisawa},\altaffilmark{3}
A. \textsc{Kubota}, \altaffilmark{4}
M. \textsc{Miyamoto}, \altaffilmark{5}
L. \textsc{Winter}, \altaffilmark{6}
}
\author{
Y. \textsc{Ueda},\altaffilmark{7}
N. \textsc{Isobe},\altaffilmark{4}
G. \textsc{Dewangan},\altaffilmark{8}
C. \textsc{Done},\altaffilmark{9}
R. E. \textsc{Griffiths},\altaffilmark{8}
Y. \textsc{Haba},\altaffilmark{10}
}
\author{
M. \textsc{Kokubun},\altaffilmark{2}
J. \textsc{Kotoku},\altaffilmark{11}
K. \textsc{Makishima},\altaffilmark{2,4}
K. \textsc{Matsushita},\altaffilmark{5}
R. F. \textsc{Mushotzky},\altaffilmark{6} 
}
\author{
M. \textsc{Namiki},\altaffilmark{12}
R. \textsc{Petre},\altaffilmark{6}
H. \textsc{Takahashi},\altaffilmark{1}
T. \textsc{Tamagawa},\altaffilmark{4}
and
Y. \textsc{Terashima}
\thanks{present address: 
Department of Physics, Ehime University, 
Bunkyo-cho, Matsuyama, Ehime 790-8577
},\altaffilmark{3}
}
\altaffiltext{1}{
Department of Physics, Hiroshima University, \\
1-3-1 Kagamiyama, Higashi-Hiroshima, Hiroshima 739-8526
}
\email{mizuno@hepl.hiroshima-u.ac.jp}
\altaffiltext{2}{
Department of Physics, University of Tokyo, 
7-3-1 Hongo, Bunkyo-ku, Tokyo 113-0033
}
\altaffiltext{3}{
Institute of Space and Astronautical Science, 
Japan Aerospace Exploration Agency, \\
3-1-1 Yoshinodai, Sagamihara, Kanagawa 229-8510
}
\altaffiltext{4}{
Institute of Physical and Chemical Research (RIKEN), 
2-1 Hirosawa, Wako, Saitama 351-0198 
}
\altaffiltext{5}{
Department of Physics, Tokyo University of Science, 
1-3 Kagurazaka, Shinjuku-ku, Tokyo 162-8601
}
\altaffiltext{6}{
Exploration of the Universe Division,
NASA Goddard Space Flight Center, Greenbelt, MD 20771, USA
}
\altaffiltext{7}{
Department of Physics, Kyoto University, Sakyo-ku, Kyoto 606-8502
}
\altaffiltext{8}{
Department of Physics, Carnegie Mellon University, 
5000 Forbes Avenue, Pittsburgh, PA 15213, USA
}
\altaffiltext{9}{
Department of Physics, University of Durham, 
South Road, Durham, DH1 3LE, UK
}
\altaffiltext{10}{
Department of Physics, Nagoya University, Furo-cho,
Chikusa, Nagoya 464-8602
}
\altaffiltext{11}{
Department of Physics, Tokyo Institute of Technology, 
2-12-1, Meguro-ku, Ohokayama, Tokyo 152-8551
}
\altaffiltext{12}{
Department of Earth and Space Science, Osaka University, \\
1-1 Machikaneyama-cho, Toyonaka, Osaka 560-0043
}

%


\maketitle

\begin{abstract}
Two ultraluminous X-ray sources (ULXs)
in the nearby  Sb galaxy NGC~1313, named X-1 and X-2, 
were observed with Suzaku on 2005 September 15.
During the observation for a net exposure of 28~ks 
(but over a gross time span of 90~ks), 
both objects varied in intensity by about 50~\%.
The 0.4--10~keV X-ray luminosity of X-1 and X-2 was measured as
$2.5 \times 10^{40}~{\rm erg~s^{-1}}$ and 
$5.8 \times 10^{39}~{\rm erg~s^{-1}}$, respectively,
with the former the highest ever reported for this ULX.
The spectrum of X-1 can be explained by a sum
of a strong and variable power-law component
with a high energy cutoff,
and a stable multicolor blackbody  
with an innermost disk temperature of $\sim 0.2$ keV.
These results suggest 
that X-1 was in a ``very high'' state,
where the disk emission is strongly Comptonized.
The absorber within NGC~1313 toward X-1 is
suggested to have a subsolar oxygen abundance.
The spectrum of X-2 is best represented, 
in its  fainter phase,
by a multicolor blackbody model
with the innermost disk temperature of 1.2--1.3~keV,
and becomes flatter as the source becomes brighter.
Hence X-2  is interpreted to be in a slim-disk state.
These results suggest that the two ULXs have 
black hole masses of a few tens to a few hundreds solar masses.

\end{abstract}

\section{Introduction}

Ultraluminous X-ray sources, or ULXs, are pointlike
X-ray sources with bolometric luminosities in excess of
$3 \times 10^{39}~{\rm erg~s^{-1}}$.
They were first discovered using the Einstein Observatory 
(for a review of the first decade of ULX discoveries,
see \cite{Fabbiano1989}),
and have often been hypothesized 
to be massive accreting black-hole binaries (BHBs)
because of their high luminosity and time variability.
Assuming a simple geometry, an accreting mass of
30--100~\MO\ is required so that the observed luminosities
do not exceed the Eddingtion limit,
although the firm evidence for this interpretation had been lacking
for almost 20 years.

A breakthrough in the study of ULXs 
was brought about by observations with the ASCA satellite:
more than a dozen ULXs were studied spectroscopically
in the 0.5--10~keV energy range 
for the first time (e.g.,
\cite{Okada1998}; \cite{Colbert1999}; 
\cite{Makishima2000}; \cite{Mizuno2000}).
Spectra of the majority of the ASCA sample have been well modeled
by a so-called multi-color disk model (MCD model: \cite{Mitsuda1984}).
Spectral transitions between the MCD-type state and the power-law type
state, often seen in
Galactic BHBs, have also been found in some ULXs
(\cite{LaParola2001}; \cite{Kubota2001b}).
These spectral characteristics generally reinforce the black hole scenario
with masses of 30--100~\MO\ for ULXs, 
although the apparent properties of high disk temperatures
(innermost disk temperature $T_{\mathrm{in}}$ within the range 1.0--1.8~keV;
\cite{Makishima2000})
and the change of innermost disk radius
(\cite{Mizuno2001}) 
both distinguish
them from simple scaled-up versions of stellar-mass BHBs.

In the early 2000s, further advancements were provided from observations made
with the  Chandra and XMM-Newton Observatories.
More than 150 off-nucleus sources,
with luminosities exceeding $10^{39}~{\rm erg~s^{-1}}$,
have been observed using Chandra,
and most of them have been determined to be pointlike 
at $0.\hspace{-2pt}^{"}5$ angular resolution (\cite{Swartz2004}).
Some luminous ULXs were found to be associated
with recent star formation activities,
suggesting that they originate in young and short-lived systems.
High-quality spectra of $\sim$ 30 ULXs,
obtained using XMM,
were in many cases dominated by a power-law like component,
often accompanied by  soft excess below 1~keV
(e.g., \cite{Foschini2002}; \cite{Feng2005}; \cite{Winter2005}).
This soft component could be interpreted as
emission from a cool accretion disk ($T_{\mathrm{in}}$=0.1--0.3~keV) 
around a massive black hole of $\sim 10^{3}~\MO$
(e.g., \cite{Miller2003}; \cite{Miller2004}),
although this interpretation is still controversial
(e.g., \cite{Dewangan2005}; \cite{Stobbart2006};
\cite{Goad2006}).

Another important clue to the nature of ULXs
has been provided through the study of
Galactic BHBs.
Besides the established two spectral states of BHBs, namely,
the low/hard state with a dominant power-law continuum
and the high/soft state dominated by an MCD emission
(e.g., \cite{McClintock2003}),
two novel states have been identified
mainly with RXTE observations of several BHBs
(e.g., \cite{Kubota2001a}; \cite{Kobayashi2003}; \cite{Kubota2004b}).
One is a so-called very high state,
known since the 1990s (e.g., \cite{Miyamoto1991}).
As a source gets more luminous than in the soft state and enters this
new state,
its spectrum again becomes power-law like,
presumably due to Comptonization by a hot optically thick plasmas,
with the MCD component becoming weaker and sometimes cooler.
By re-analyzing the ASCA data of
IC~342 source~1,
\citet{Kubota2002} showed that the observed power-law type spectrum
of this  ULX is suggestive of the very high state.
The fourth state,
characterized by an MCD-like but hotter spectra,
is seen when a source is very luminous,
close to or even exceeding the Eddington limit.
As argued by \citet{Kubota2004b} through the observation
of Galactic BHB XTE J1550--564,
this state may correspond to the theoretically predicted 
"slim disk" solution
(e.g., \cite{Abramowicz1988}; \cite{Szuszkiewicz1996}; \cite{Watarai2000}),
in which an optically-thick disk with a moderate geometrical 
thickness is formed and advective cooling becomes important.
The slim-disk interpretation successfully explains
the high disk temperature and the changes of the disk radius
observed from the most luminous class of ULXs
(e.g., \cite{Mizuno2001}; \cite{Ebisawa2003}).

Despite the great progress described above, 
our understanding of ULXs is still far from satisfactory.
In particular, it has been difficult to unambiguously
distinguish different modelings of their spectra,
and to understand how the spectrum changes as the source varies.
In the present paper,
we report our new Suzaku results on
two ULXs (X-1 and X-2) in the nearby spiral galaxy NGC~1313.
We suggest that X-1 was in a state which is very similar
to the very high state of Galactic BHBs, while X-2
was in another state which may correspond to the 
slim disk solution.
Our new results also include possible indication of
sub-solar oxygen abundance in the X-ray absorption toward X-1.


\section{Instrumentation and Observation}

The 5th Japanese X-ray satellite Suzaku 
(\cite{Mitsuda2006}) has several important
instrumental properties which are suited to ULX studies.
The X-ray Imaging Spectrometer (XIS; \cite{Koyama2006}),
combined with the X-ray Telescope (XRT; \cite{Serlemitsos2006}),
is our primary observing tool.
Among the four XIS cameras, one (XIS1)
utilizes a back-side illuminated CCD chip (BI chip),
and has an unprecedented sensitivity and energy resolution
in the energy band below 1~keV.
The other three XIS cameras, with front-side illuminated chips (FI chips),
have an even better energy resolution,
together with very low and stable background toward the hardest spectral end.
As a result, the XIS plus XRT combination
covers a very wide energy range, 0.2--12~keV,
with a much flatter response than was achieved with previous missions.
This in turn  allows us to accurately constrain the continuum shape of ULXs.
In addition, the Hard X-ray Detector 
(HXD; \cite{Takahashi2006}; \cite{Kokubun2006})
has unprecedented sensitivity
in the energy range above 10 keV,
and may detect
the hard X-ray emission from the brightest ULXs.

NGC~1313 is a nearby face-on, late-type Sb galaxy at a distance of
3.7~Mpc (\cite{Tully1988}).
Although a more recent estimate by \citet{Mendez2002} gives 4.13Mpc, 
we adopt the value by \citet{Tully1988}
for consistency with other works on ULXs in this galaxy 
(e.g., \cite{Miller2003}; \cite{Dewangan2005}; \cite{Stobbart2006}).
If 4.13 Mpc is adopted,
all the luminosities quoted in the present work increases by 20\%.
The low Galactic line-of-sight absorption towards NGC~1313
($N_{\mathrm{H}}=0.35 \times 10^{21}~{\rm cm^{-2}}$)
allows us to study low energy spectral properties of sources in it.
According to previous X-ray observations of NGC~1313
using the Einstein Observatory and ROSAT
(\cite{Fabbiano1987}; \cite{Colbert1995}),
its X-ray emission is dominated by three extremely luminous 
pointlike sources of X-ray luminosity 
$L_{\mathrm{X}} \sim 10^{39}~{\rm erg~s^{-1}}$ each.
One of them is the unusually luminous X-ray supernova, namely SN~1978K.
The other two are typical ULXs,
which we call X-1 and X-2 after \citet{Colbert1995};
the former is located close ($\sim 45^{''}$) to the galaxy nucleus.

The two ULXs have already been studied extensively with
ASCA and XMM-Newton
(e.g., \cite{Makishima2000}; \cite{Mizuno2001}; \cite{Miller2003};
\cite{Dewangan2005}; \cite{Stobbart2006}; \cite{Feng2006}).
ASCA observations found moderate amplitude
long-term variability of these ULXs.
While the spectral behavior of X-1 was rather complex,
the ASCA spectrum of X-2 was represented by a hot (1.0--1.5~keV)
MCD model (\cite{Mizuno2000}; \cite{Mizuno2001}).
The XMM spectra of both ULXs in 2000
can be represented by a power-law with soft excess,
which \citet{Miller2003} interpreted as MCD emission
from a cool accretion disk around a very massive
($\sim 1000~\MO$) BH.
Optical observations of X-2 indicate
that the source is a high-mass X-ray binary system
(\cite{Zampieri2004}; \cite{Mucciarelli2005}).

\begin{figure}
\begin{center}
\FigureFile(160mm,160mm){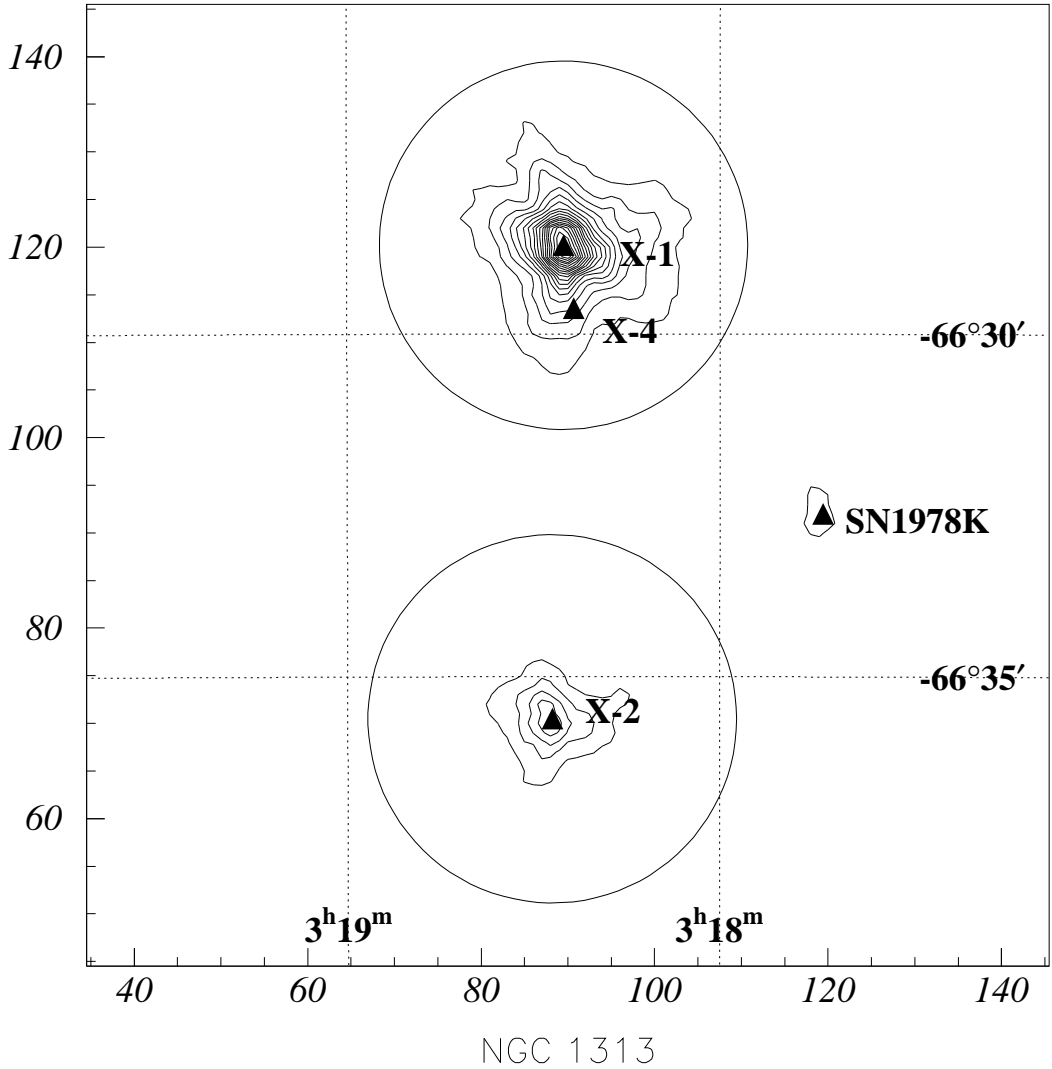}
\end{center}
\caption{
An XIS0 image of the NGC~1313 region in the 0.4--10~keV band,
where we can see X-1, X-2 and SN~1978K as local peaks in the X-ray contours.
The image was smoothed using a gaussian distribution with
$\sigma=0.\hspace{-2pt}^{'}1$.
Triangles indicate the positions of strong X-ray sources detected
in the ROSAT HRI observation
(\cite{Schlegel2000}), shifted by $0.\hspace{-2pt}^{'}65$ which is
within the pointing accuracy of the Suzaku satellite.
Also shown are the accumulation regions for source spectra.
Note that source names are after \citet{Colbert1995}.
}
\label{fig:NGC1313_sky.eps}
\end{figure}

We conducted a Suzaku observation of NGC~1313 on 2005 October 15,
with the XRT optical axis aimed at the middle point between X-1 and X-2.
The data were screened based on the following standard criteria:
a) The time elapsed after a passage through the South Atlantic Anomaly
is longer than 256 seconds,
b) The object is at least $5^{\circ}$ and $20^{\circ}$
above the rim of the Earth during night and day, respectively, and
c) The geomagnetic cutoff rigidity is greater than 6~GV.
After these data screenings,
the net exposure with the XIS became 27.8~ks.
The resulting XIS image of the galaxy is shown in
figure~\ref{fig:NGC1313_sky.eps}.
The spectrum of the HXD, after subtracting 
the instrumental background model provided by
the instrument team, is consistent with that of the cosmic
X-ray background (\cite{Boldt1987}),
and no significant hard X-ray emission above 10~keV was detected.

\section{Data Analysis}

\subsection{Light Curves of two ULXs}

\begin{figure}
\begin{center}
\FigureFile(160mm,106mm){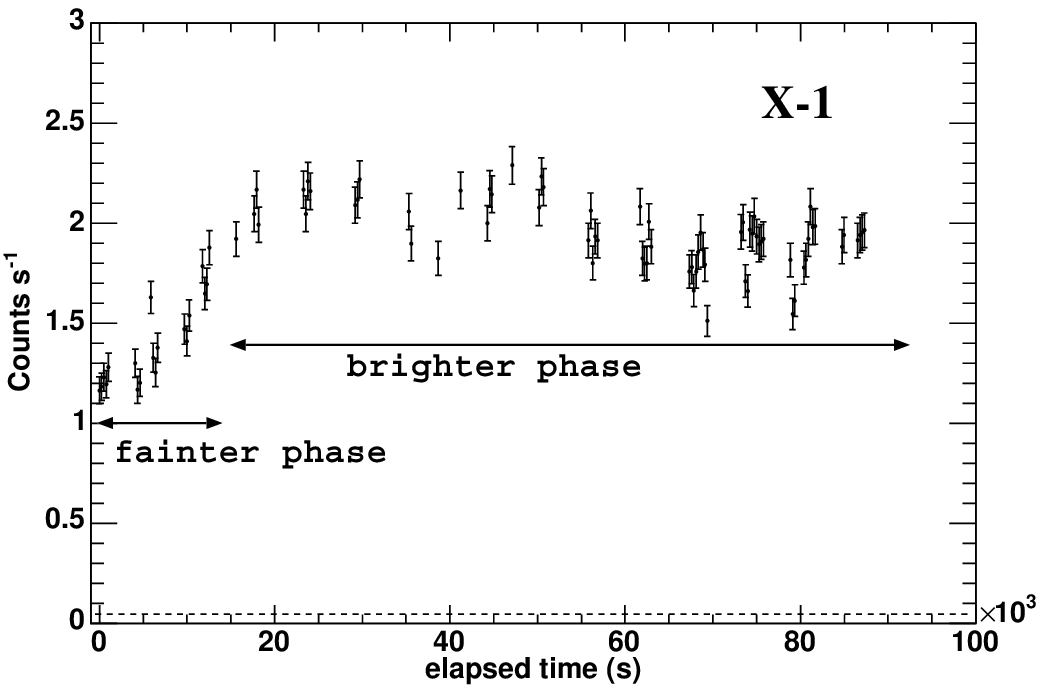}
\FigureFile(160mm,106mm){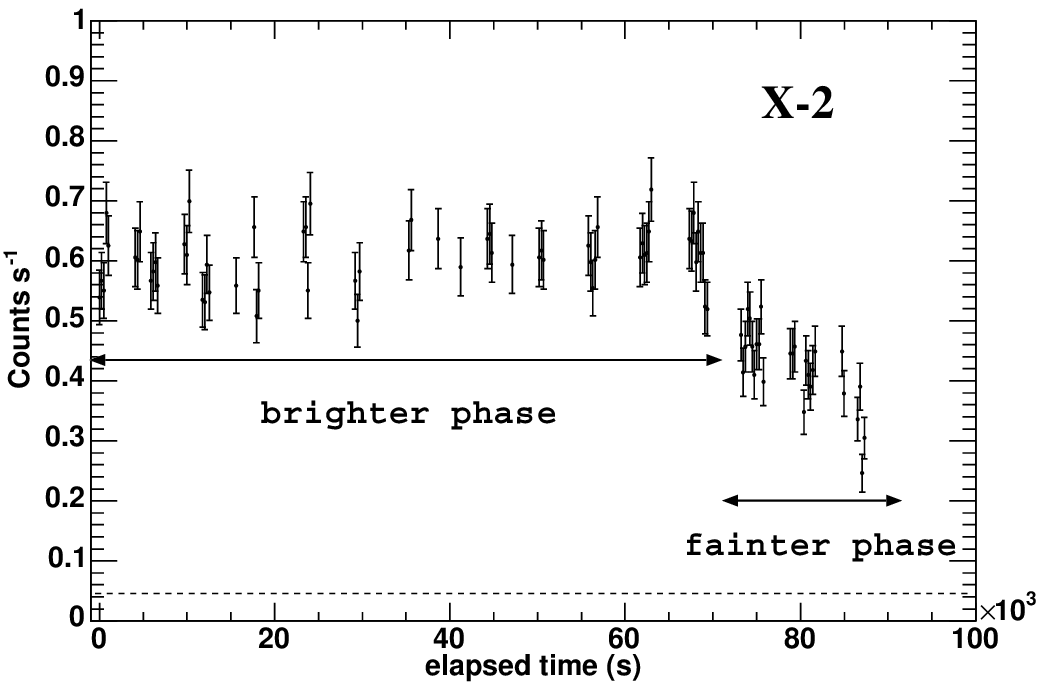}
\end{center}
\caption{
Suzaku FI light curves for X-1 (top panel) and X-2 (bottom panel)
in the 0.4--10~keV energy range with 256~s binning.
Background is included and the background level is indicated
by dotted lines. Also shown are the time intervals 
used to study the
short-term spectral variability.
}\label{fig:lightCurve}
\end{figure}

The 0.4--10~keV light curves of NGC1313 X-1 and X-2, extracted from the
circular regions of $3^{'}$ radius,
are shown in figure~\ref{fig:lightCurve}.
Thus, the two ULXs both varied by $\sim$50\%, 
with a flux increase  in  X-1 and a decrease  in X-2.
The variations are intrinsic to the sources 
rather than being instrumental artifacts (e.g., vignetting of the mirror),
since we used the data after the satellite position was well stabilized,
and because both objects were placed rather close
($\sim 3.\hspace{-2pt}^{'}5$) to the optical axis of the telescope
compared to the XIS field-of-view ($18^{'} \times 18^{'}$).

To investigate the change of spectral shapes
in a model independent way,
we subtracted background 
extracted from a source free region of the same observation,
and calculated the ratio of the spectrum 
of the fainter phase to that of the brighter phase.
The observed ratios, shown in  figure~\ref{fig:specRatio}, 
reveal intensity-correlated spectral changes in both ULXs. 
In X-1, the ratio stays at $\sim 1$ below 1~keV, 
and decreases toward higher energies, 
reaching $\sim 0.6$ at 3~keV and then flattening.
The positive correlation between the intensity and spectral hardness,
observed from X-1 in this particular observation, 
is in the opposite sense as found on other occasions 
(e.g., \cite{Dewangan2005}; \cite{Feng2006}).
The X-2 spectrum softens as the source gets dimmer,
implying also a positive intensity vs. hardness correlation,
but the spectral change is particularly prominent
in energies above $\sim 3$ keV.

In the following sections, we first analyze the time-averaged spectra
to grasp rough spectral information (\S~\ref{sec:time-average}),
and then study the spectral variability in detail
(\S~\ref{sec:variability}).

\subsection{Time-averaged Spectra}
\label{sec:time-average}

We first calculated spectral ratios of the two ULXs
to the Crab Nebula (Crab ratio).
Since the Crab Nebula is known to have a stable, simple power-law
X-ray spectrum of the photon index $\Gamma \sim 2.1$
(\cite{Toor1974}), 
the Crab ratios can be used to roughly examine 
spectral shapes of various X-ray sources
in the form close to $\nu F\nu$ representation,
without using a detector response.
As shown in figure~\ref{fig:crabRatio},
the X-1 spectrum has a power-law like shape up to 3~keV,
and then falls off above 5~keV.
The X-2 spectrum shows a similar shape, 
but is more convex than that of X-1.
Thus, neither spectrum can be well modeled by a single power-law.

We then fitted the time-averaged spectra
by xspec of version 11.3
using the response matrix files as of 2006 February 13,
and auxiliary response files generated by the program
xissimarfgen as of 2006 April 24.
Since the two ULXs exhibited clear spectral changes
during the observation, 
here we aim at roughly evaluating spectral properties 
rather than quantifying the spectral parameters.
We employed the MCD plus power-law model
(hereafter called "MCD+PL" model), 
a standard model widely used to describe the spectra of BHBs and ULXs.
The data from the three FI cameras were 
summed together to improve the statistics.
Contamination on the XIS by out-gassing from the satellite
(\cite{Koyama2006})
and the consequent decrease in the low-energy efficiency were
taken into account using the xispcoab model
\footnote{
http://heasarc.gsfc.nasa.gov/docs/xanadu/xspec/models/xisabs.html
}
developed by the XIS team;
this empirically models the positional dependence and time evolution
of the contaminant thickness of each XIS sensor
based on calibration observations.
Two free parameters of the model, 
namely the date of observation
and the target offset angle from the XIS nominal position,
were fixed at 63 and $3.\hspace{-2pt}^{'}5$, respectively;
the model gave a column density of 
1.0--2.9$\times 10^{18}~{\rm cm^{-2}}$ and
0.18--0.48$\times 10^{18}~{\rm cm^{-2}}$ for
carbon and oxygen in the contaminant, respectively.
The line-of-sight absorption outside the satellite
was taken into account using the ``wabs'' model (\cite{Morrison1983}),
with hydrogen column density as a free parameter.
Since a small gain jump at the Si K edge is not yet
properly handled in the current pipeline processing, 
data with PI channels 500--504  (corresponding to 1.82--1.84~keV) 
are ignored in the fitting.

As shown by figure~\ref{fig:spectra},
the model approximately reproduces the data,
although the fit is rather poor with $\chi^2/\nu = 442.6/282$, 
mainly because of residuals in the 0.4--0.6~keV band and around 2~keV.
Since the shape of the residuals around 2~keV is different 
between the FI and BI spectra, 
this is likely to be instrumental due to calibration uncertainties.
We therefore exclude data in the 1.5--2.3~keV band
(which covers the 1.82--1.84 keV range already excluded)
in the following analysis of the X-1 spectrum.
The low-energy residuals are prominent below the neutral O-edge absorption, 
and is suggestive of an incorrect amount of oxygen in the spectral model.
We therefore exclude, for the moment,
the X-1 data below 0.6~keV as well,
and investigate this issue in \S~\ref{sec:lowEnergyResidual}.
Neglecting the data in these two energy regions gives a marginally 
acceptable  fit to the X-1 spectrum with $\chi^2/\nu = 256.9/218$.
The X-2 spectrum, in contrast, 
is reproduced successfully by the  MCD+PL model with
$\chi^2/\nu = 152.2/142$, 
probably because of the limited photon statistics.
We thus utilize the full energy band of the Suzaku XRT+XIS
(except the  1.82--1.84 keV range)
in the following spectral analysis of X-2.

The best fit model parameters describing the time averaged 
X-1 and X-2 spectra  are tabulated in  table~\ref{tab:time-av},
and contributions of the individual model components
are shown in  the insets to figure~\ref{fig:spectra}.
The power-law component and the MCD component
contribute almost equally to the X-1 spectrum,
whereas the MCD component dominates the X-2 spectrum.
These properties agree with the inference from the Crab ratio 
(figure~\ref{fig:crabRatio}).
Assuming the source distance of
3.7~Mpc (\cite{Tully1988})  and isotropic radiation,
the 0.4--10~keV source luminosity is 
$2.5 \times 10^{40}~{\rm erg~s^{-1}}$ and
$5.8 \times 10^{39}~{\rm erg~s^{-1}}$ for X-1 and X-2,
respectively, after correction for the absorption
inside and outside the spacecraft.
Even allowing for possible calibration uncertainties by $\sim 10\%$,
the flux of X-1 is the highest among those ever reported for this source
including all archival XMM-Newton data available as of June 2006, 
as we confirmed through our own analysis;
more than a factor of 4 higher 
than that of the XMM-Newton 2000 observation.
The flux of X-2 is between those of the 1993 and 1995 ASCA observations,
and a factor of 3 higher than that measured in the XMM-Newton observation
in 2000 (Mizuno et al. 2001; \cite{Miller2003}).

\subsection{Spectral Variability of Each Source}
\label{sec:variability}

To quantify the spectral changes (figure~\ref{fig:specRatio})
associated with the intensity variations, 
we then divided the data into two time regions shown
in figure~\ref{fig:lightCurve},
and derived ``brighter phase'' and ``fainter phase'' spectra
from the two sources. We then fitted them with
several models widely used to represent the spectra of ULXs:
the MCD+PL model, so called $p$-free disk model,
and the MCD plus cutoff power-law model (hereafter MCD+cutoff-PL).
The MCD+PL model is the most commonly used 
description of BHB spectra, mainly those in the soft state.
The $p$-free disk model is a modified MCD model,
where the disk temperature profile as a function of radius $r$ is given as
$T(r) = T_{\mathrm{in}}(r/r_{\mathrm{in}})^{-p}$ 
with $p$ being a positive free parameter.
This model was originally developed by \citet{Mineshige1994}
to validate the standard accretion disk prediction on the 
temperature profile ($p=0.75$). The model was later utilized
to represent the spectra of BHBs and ULXs when the objects are
thought to be in the slim disk state (e.g., \cite{Kubota2004b}),
since the spectrum emergent from a slim disk is theoretically predicted 
to become flatter, and be approximated by the $p$-free disk model 
with $p$ decreasing to $\sim 0.5$ (\cite{Watarai2000}).
A power-law type hard spectrum with soft excess below 1~keV
has been widely observed in ULXs including those in NGC~1313
(\cite{Miller2003}).
Since the apparently power-law like continuum often exhibits,
in close inspection, a high-energy turn over
(e.g., \cite{Kubota2002}; \cite{Dewangan2005}; \cite{Stobbart2006}),
we also employed the MCD+cutoff-PL model
in which the MCD component represents the
excess in the soft band and the cutoff-PL represents the
curved spectrum in the high energy band.
The role of the MCD component in this model is 
different from that in the first one (MCD+PL), 
in which the MCD component has a high temperature
and represent the high-energy spectral curvature.

All these three models are moderately successful on both
the fainter-phase and brighter-phase spectra of X-1,
yielding the best fit parameters as summarized in table~\ref{tab:time-sort}.
The MCD and $p$-free disk models allow us to evaluate the
innermost disk radius $R_{\mathrm{in}}$. 
After \citet{Makishima2000},
we calculated this parameter as 
$R_{\mathrm{in}} = \xi \kappa^{2} \sqrt{N/\cos i} ~(D/{\rm 10~kpc})$,
where $N$ is the model normalization,
$i$ is the disk inclination angle,
and $D$ is the distance to the source;
$\xi=0.41$ and $\kappa=1.7$ are correction factors described in
\citet{Kubota1998} and \citet{Shimura1995}.
The best-fit spectrum of each phase  is given in figure~\ref{fig:spectra_x1}.
Physical meanings of the obtained fits are discussed in
\S~\ref{sec:discuss_x1}.

We obtained similar results on the brighter-phase X-2 spectrum:
the three model are all moderately successful, with a rather
small difference in their fit goodness (table~\ref{tab:time-sort}).
The fainter-phase spectra of X-2 were reproduced by a single MCD model,
and adding a power-law component with $\Gamma$=1.0--2.5 
did not improve the fit significantly 
(less than $90\%$ confidence using the $F$-test).
In table~\ref{tab:time-sort}, 
we hence give only the single-MCD fit result
for the fainter-phase X-2 spectrum.
The spectra of X-2 in its fainter and brighter phases are
shown in figure~\ref{fig:spectra_x2} 
together with the best-fit models
(see \S~\ref{sec:discuss_x2} for the discussion 
of physical meanings of the model fits).

\subsection{Residuals at Low Energies}
\label{sec:lowEnergyResidual}
Finally, we revisit the residuals seen at low energies in the X-1 spectrum,
particularly around the oxygen absorption edge 
(at 0.53 keV in figure~\ref{fig:spectra}).
These residuals suggest 
that the model is over-predicting the edge depth.
Then, a likely origin
is an overestimation of oxygen column density in the XIS contamination model,
or an incorrect modeling of oxygen abundance 
in absorbing materials along the line of sight to the source.

To better estimate the oxygen column density in the overall absorber,
we followed the procedure of \citet{Cropper2004},
and decomposed the absorption into the following two  factors.
One is the line-of-sight Galactic absorption, 
represented by a ``wabs'' model with solar abundance ratios,
of which  the column density is fixed 
at $N_{\mathrm{H}} = 0.35 \times 10^{21}~{\rm cm^{-2}}$.
The other is  absorption inside NGC~1313 
(plus errors in the XIS contamination modeling),
expressed by  a ``tbvarabs'' absorption model (\cite{Wilms2000})
in which abundances of individual elements can be varied.
We selected the MCD+cutoff-PL model to represent the continuum,
because it can reproduce the brighter phase spectrum with
the smallest $\chi^{2}/\nu$ among the three model fits.
The elemental abundances refer to \citet{Anders1982},
which are close to those employed in the ``wabs'' model.
We refitted the data,  
leaving  the oxygen abundance  in the tbvarabs model to vary freely,
but fixing other elemental abundances at the solar values.
Indeed, the fit was improved ($\Delta \chi^{2}=-50.3$)
by lowering the  oxygen abundance to $0.51^{+0.11}_{-0.05}$ solar.
According to an $F$-test,
this improvement is significant at the 99\% confidence limit. 

We further allowed the FI and BI data to have
independent absorption parameters, to see if there are
any possible instrumental uncertainty.
Then, the FI and BI data gave consistent column densities 
of the second absorption factor as
$N_{\mathrm{H}}=(3.3\pm0.3) \times 10^{21}~{\mathrm{cm^{-2}}}$
and
$N_{\mathrm{H}}=(2.8\pm0.2) \times 10^{21}~{\mathrm{cm^{-2}}}$,
respectively,
whereas they somewhat disagreed on the circum-source oxygen abundance;
$0.31\pm0.10$ solar (FI) and  $0.68\pm0.13$ solar (BI).
Since the FI sensors have lower soft X-ray efficiency than the BI sensor, 
and may suffer from larger calibration uncertainties,
we adopt the result by the BI data, which still implies the 
subsolar oxygen abundance.

If we adopt the 0.68-solar oxygen abundance in the tbvarabs model,
the intervening oxygen column density decreases
by $0.67 \times 10^{18}~{\mathrm{cm^{-2}}}$ 
from that implied by the one-solar abundance modeling.
This decrement is significantly larger than the
oxygen column density contained in the BI sensor contaminant,
namely  $0.21 \times 10^{18}~{\rm cm^{-2}}$.
Therefore, the required reduction in the oxygen column density 
cannot be attributed to any uncertainty in the modeling 
of chemical composition and/or thickness of the XIS contaminant.
Instead, we infer that the absorber within NGC~1313 toward X-1 
has a subsolar oxygen abundance.

We performed the same analysis on X-2.
Namely, we refitted the spectrum by allowing  the oxygen abundance
in the tbvarabs model  to vary freely.
Due to the limited photon statistics, however, 
the oxygen abundance in this absorption model was only poorly constrained,
with an upper limit  of $\le 1.01$ solar.
Thus, the X-2 data are consistent with the absorber in NGC~1313 
having a sub-solar oxygen abundance,
but do not require it.

\section{Discussion}
In the Suzaku observation of NGC~1313 
spanning a gross time interval of 90 ks (but the net exposure being 28 ks),
we detected gradual intensity changes by $\sim 50$\% 
from the two ULXs, X-1 and X-2.
When time averaged,
X-1 exhibited a 0.4--10~keV luminosity of $2.5 \times 10^{40}$ erg~s$^{-1}$,
which is the highest value ever reported for this source,
while X-2 showed $5.8 \times 10^{39}$ erg~s$^{-1}$.
Thanks to the
excellent performance of the Suzaku XIS,
we obtained high-quality spectra from both ULXs,
and revealed their intensity-correlated spectral changes.
The high flux of X-1 also allowed us to obtain evidence 
of a subsolar oxygen abundance in the absorption toward it.
Below we examine spectral states of the two ULXs,
and discuss environment of the NGC~1313 X-1 system.

\subsection{Spectral State of X-1}
\label{sec:discuss_x1}
\begin{figure}
\begin{center}
\FigureFile(160mm,108mm){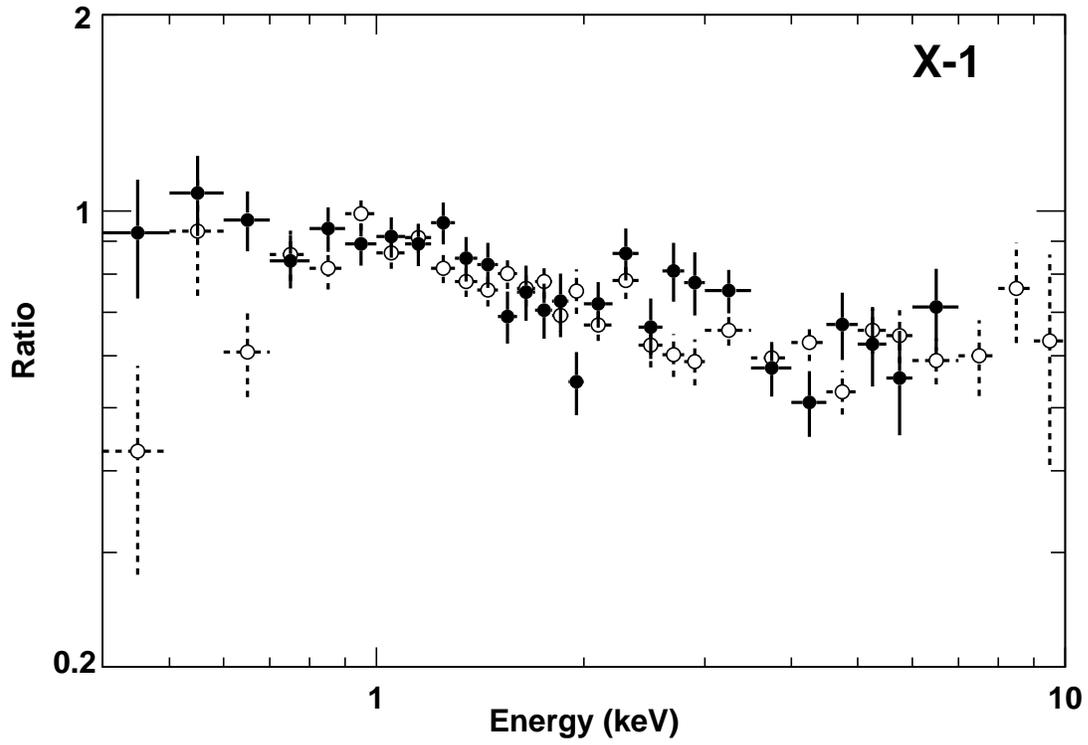}
\FigureFile(160mm,108mm){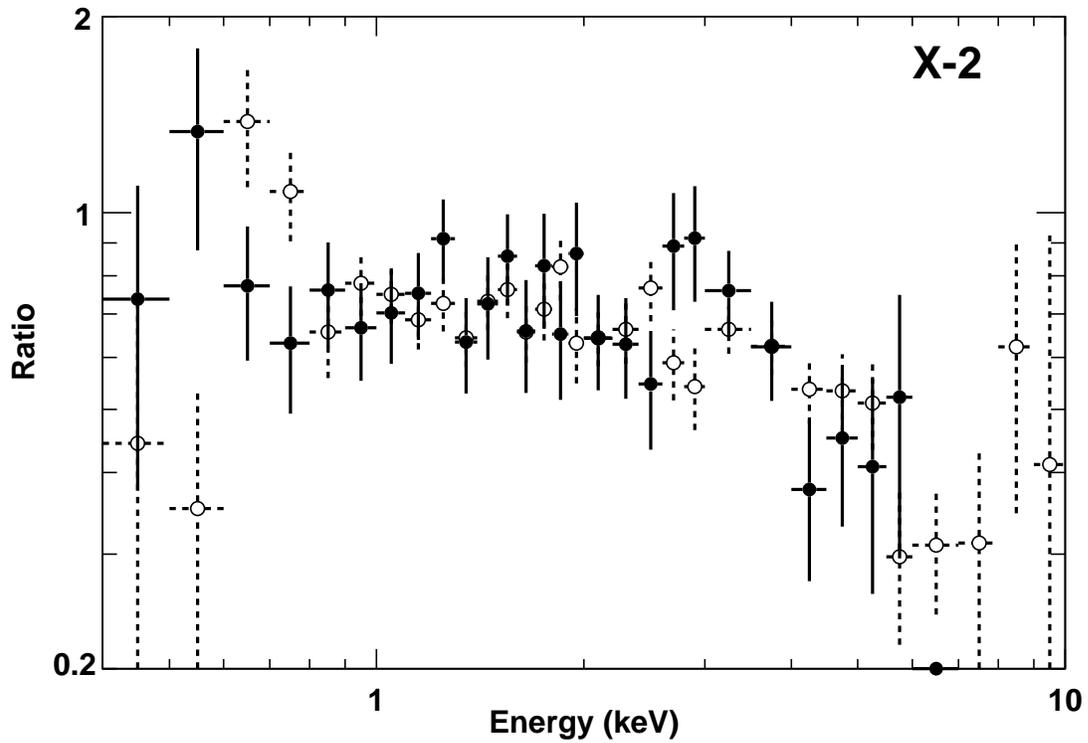}
\end{center}
\caption{
Spectral ratios of the fainter phase to the brighter phase
for X-1 (top panel) and X-2 (bottom panel),
calculated after subtracting the background. 
The ratio of BI and FI data are given as
filled circles and open circles, respectively.
}\label{fig:specRatio}
\end{figure}

\begin{figure}
\begin{center}
\FigureFile(160mm,108mm){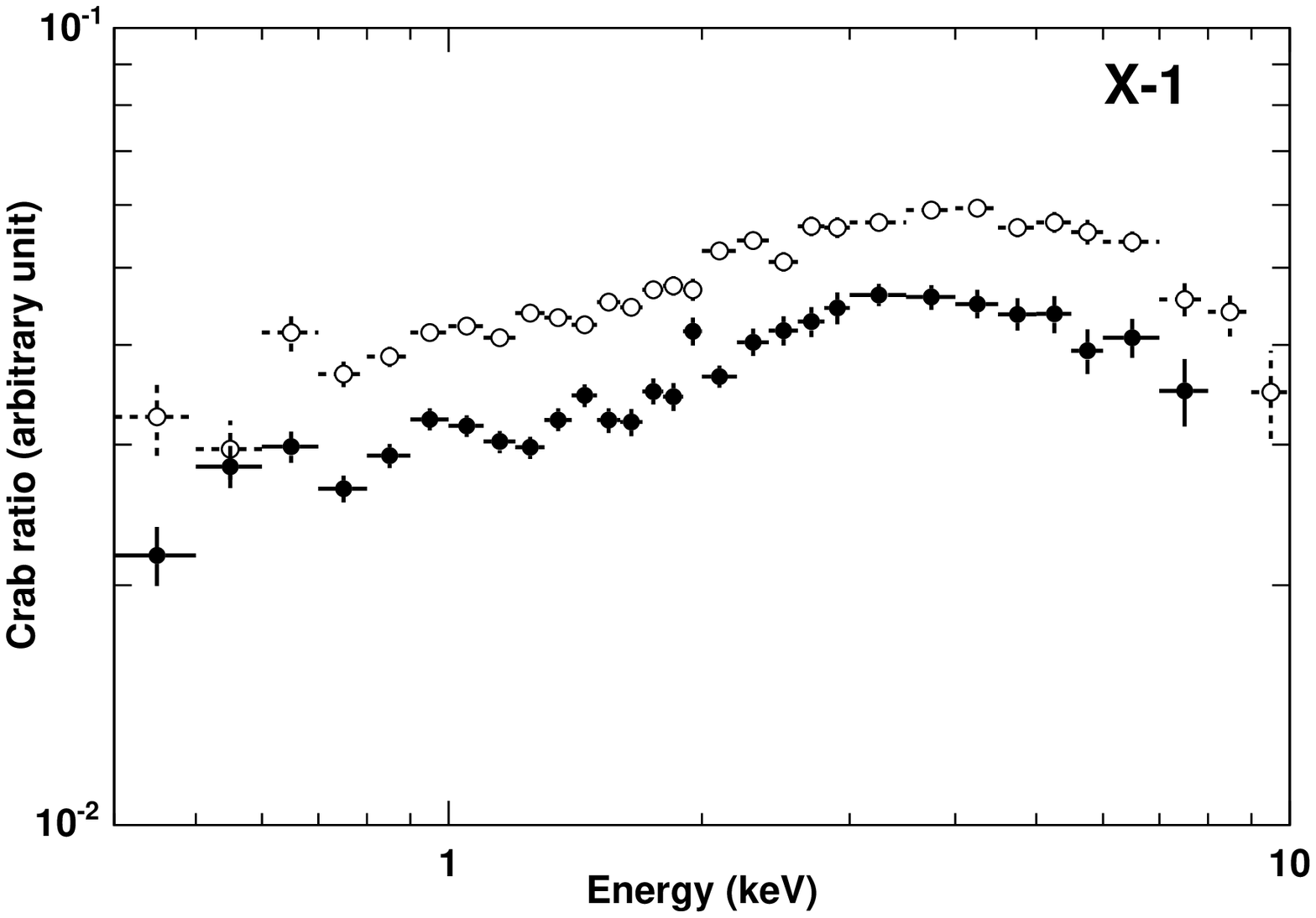}
\FigureFile(160mm,108mm){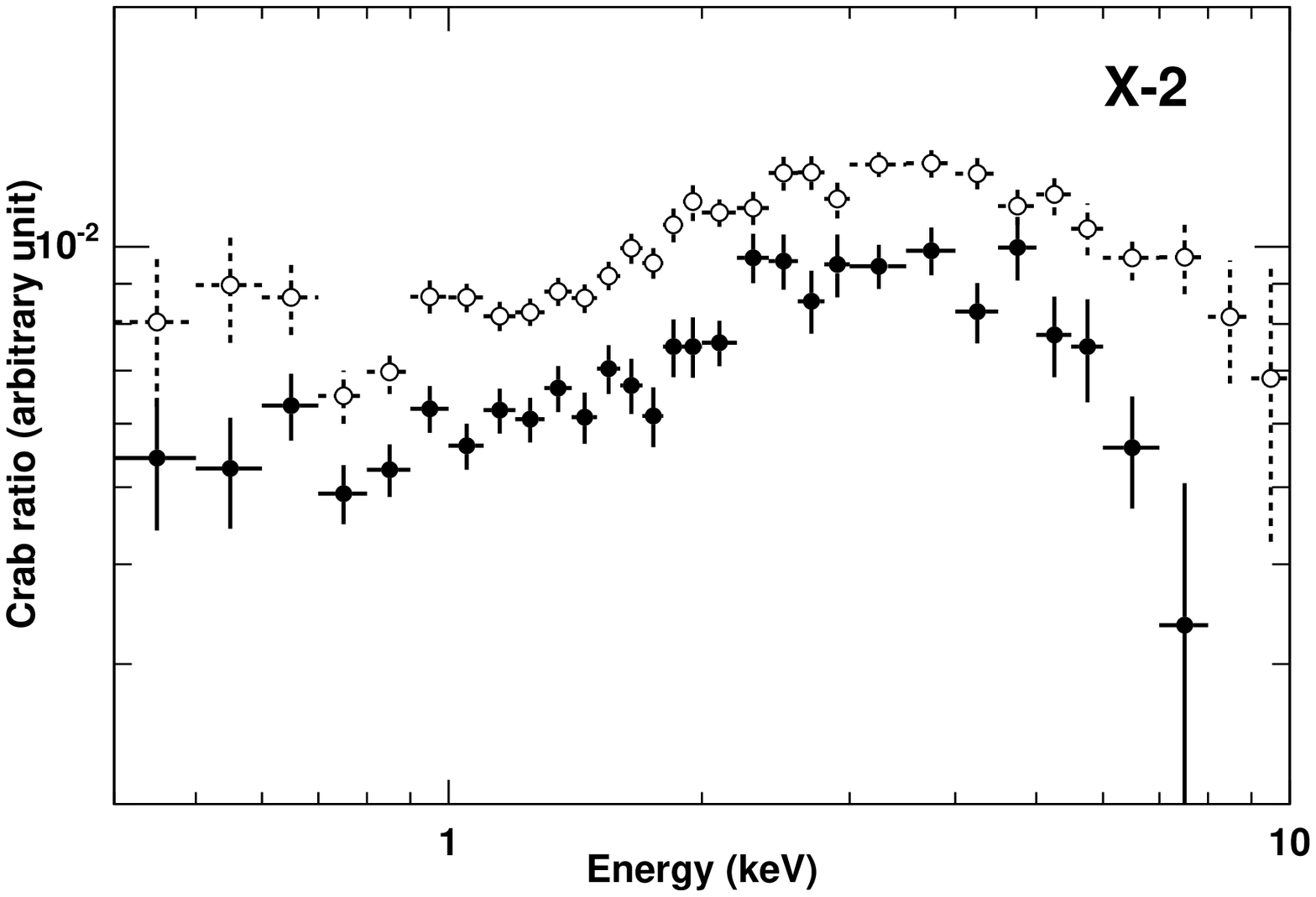}
\end{center}
\caption{
Crab ratio of the time-averaged spectra
for X-1 (top panel) and X-2 (bottom panel),
calculated after subtracting the background for two ULXs. 
Symbols are the same as those of Figure~\ref{fig:specRatio}.
FI data are shifted up for clarity.
}\label{fig:crabRatio}
\end{figure}

\begin{figure}
\begin{center}
\FigureFile(144mm,102mm){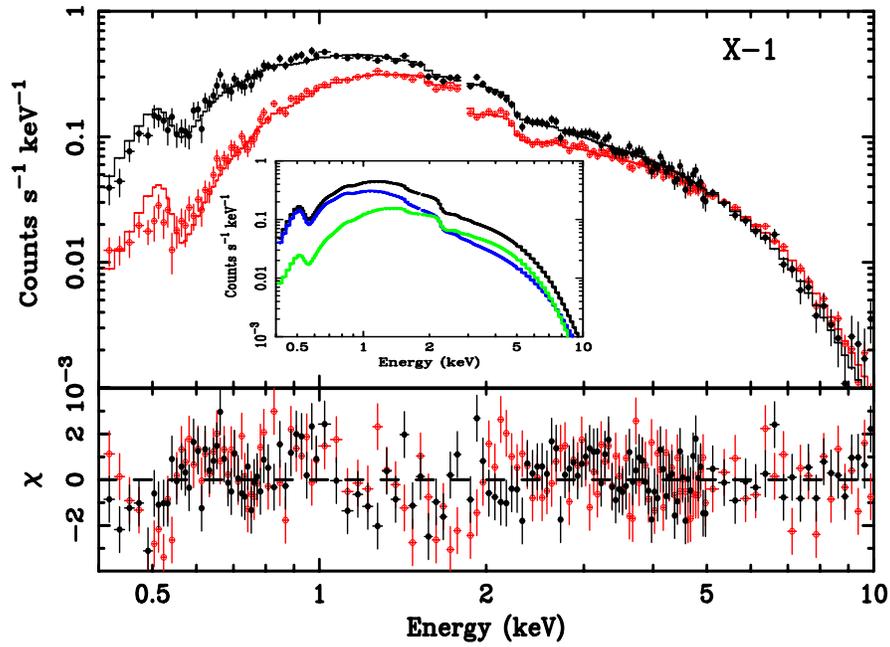}
\FigureFile(144mm,102mm){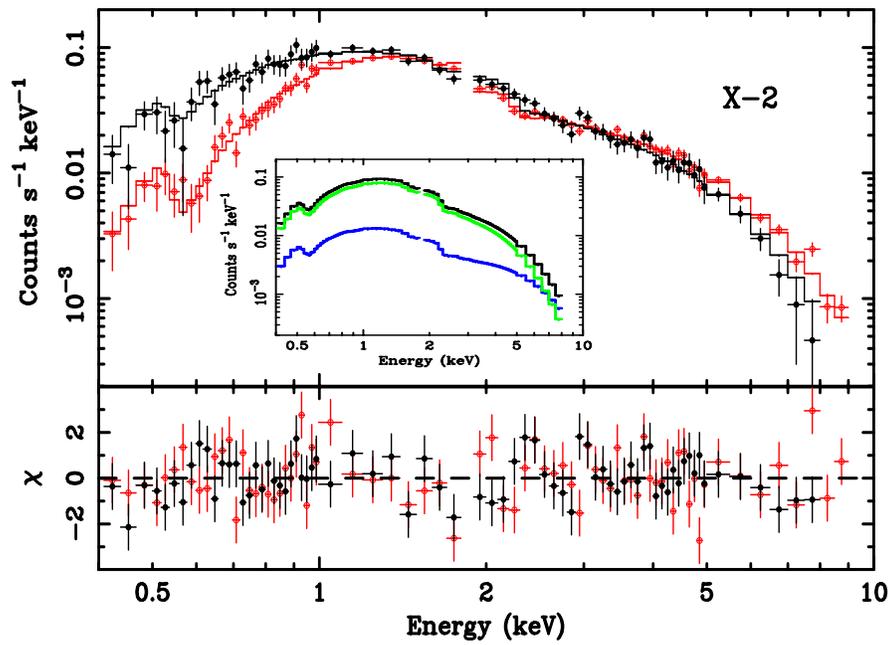}
\end{center}
\caption{
Time averaged XIS spectra of NGC~1313 X-1 (top panel) and X-2 (bottom panel)
fitted with the MCD+PL model. The histogram shows the best fit model
and crosses represent the observed spectra. 
The bottom section of each panel shows 
the fit residuals. The BI (black) and FI (red) spectra
are fitted simultaneously except for the relative normalization.
The insert shows the best-fit model (black histogram)
and contributions of each model 
(MCD and PL model shown by green and blue histogram, respectively) for BI.
The data in the 1.82--1.84~keV range are ignored
(see text).}\label{fig:spectra}
\end{figure}

\begin{figure}
\begin{center}
\FigureFile(152mm,107mm){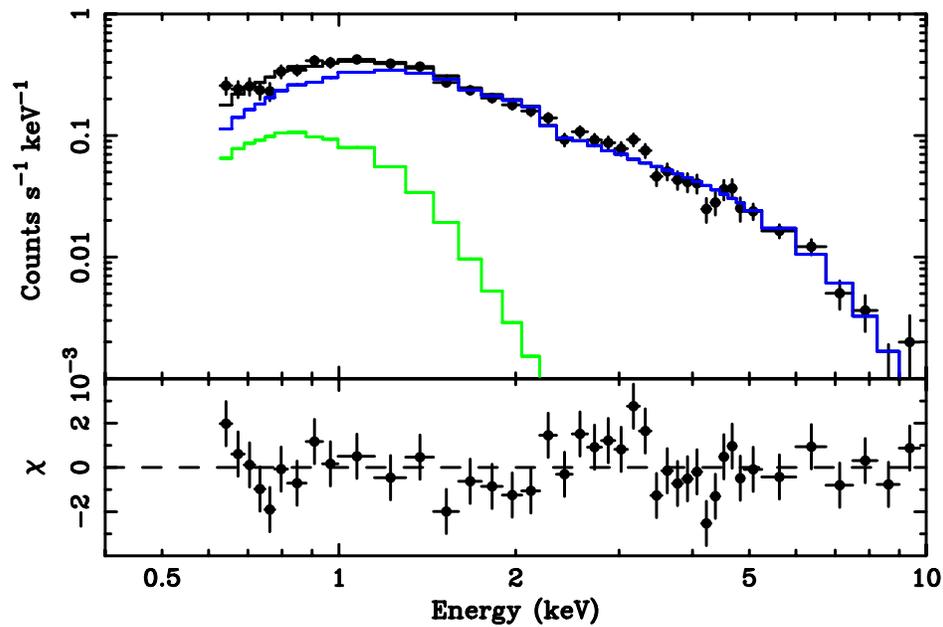}
\FigureFile(152mm,107mm){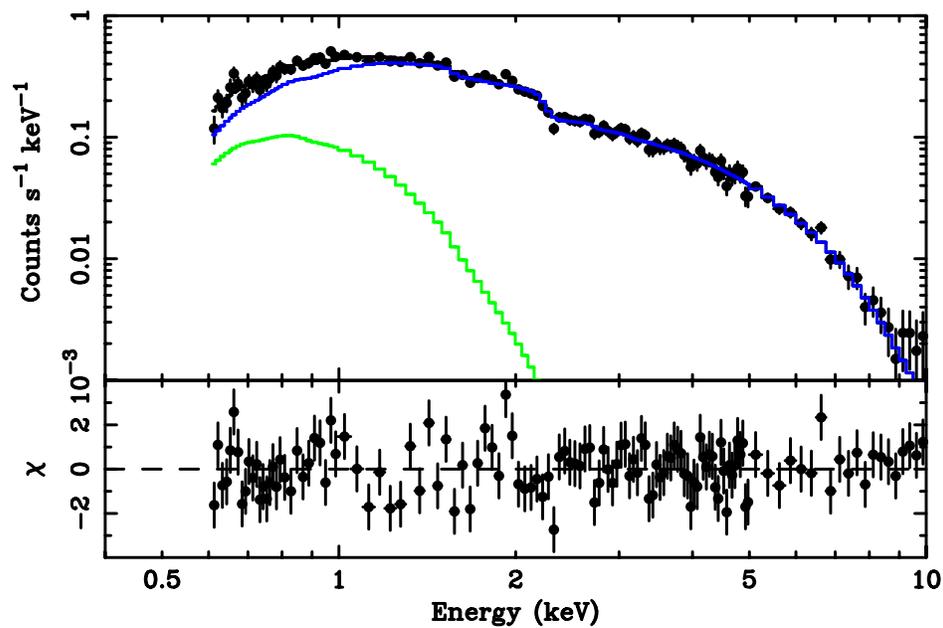}
\end{center}
\caption{
X-1 spectra for the fainter phase (top panel) 
and the brighter phase (bottom panel)
fitted with the MCD+cutoff-PL model. Symbols are the
same as those of figure~\ref{fig:spectra}.
Contribution of the MCD and cutlff-PL model are
given by green and blue histogram, respectively.
Only BI sensor data is shown for clarity.
Note that data in 1.5--2.3~keV band is not used for the fitting.
}\label{fig:spectra_x1}
\end{figure}

\begin{figure}
\begin{center}
\FigureFile(160mm,113mm){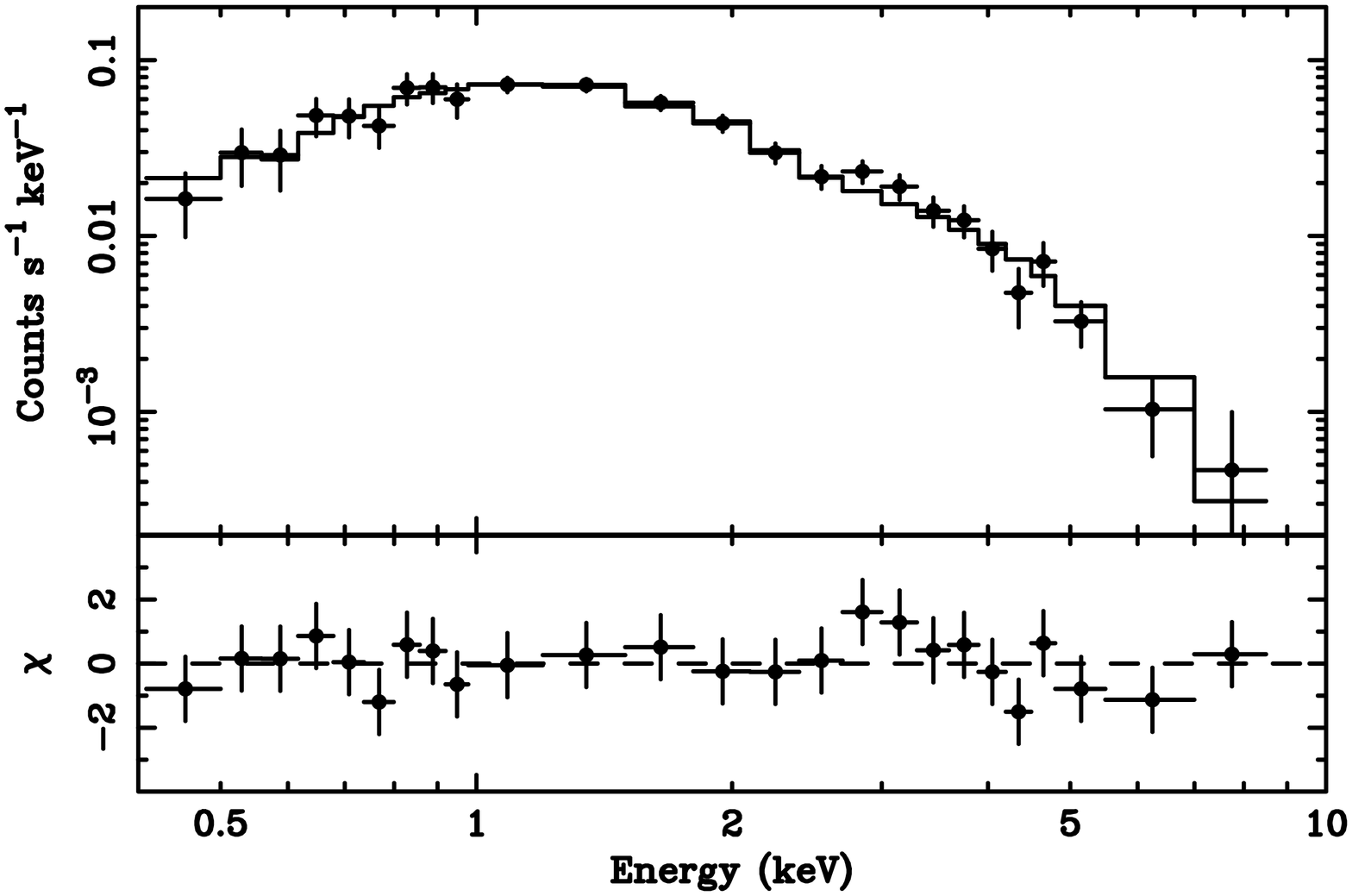}
\FigureFile(160mm,113mm){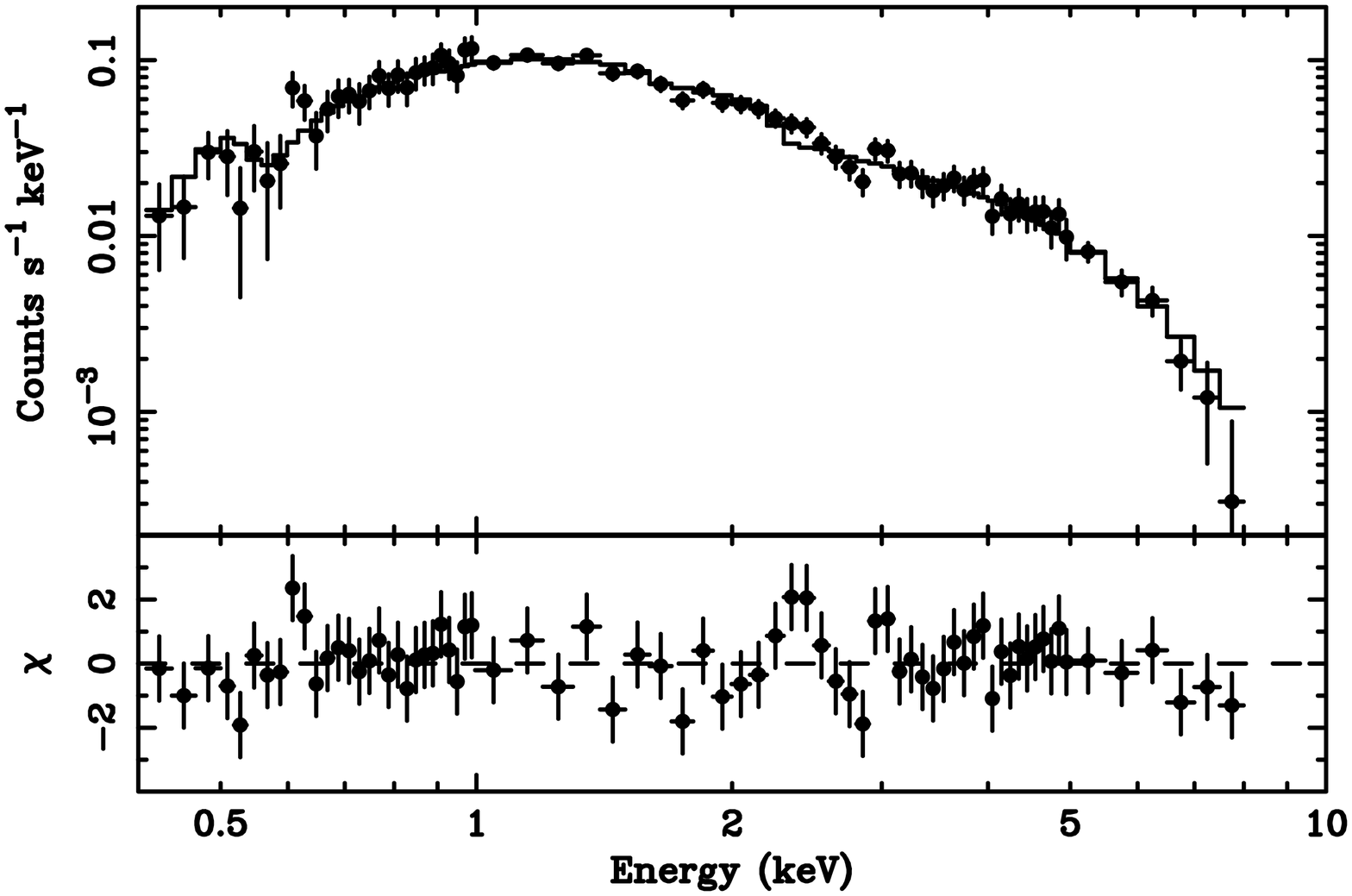}
\end{center}
\caption{
The same as figure~\ref{fig:spectra_x1}, but for X-2
instead of X-1. The fit with the MCD model alone is used
for the fainter phase spectrum (top panel) and
that with the $p$-free disk model is used
for the brighter one (bottom panel).
}\label{fig:spectra_x2}
\end{figure}

In the fainter phase, 
NGC~1313 X-1 exhibited a more or less straight continuum,
with the power-law component in the MCD+PL fit 
contributing more than 80\% (see table~\ref{tab:time-sort}).
In the brighter phase, in contrast,
the spectrum was much more convex (top panel of figure~\ref{fig:specRatio}. 
In both phases, however,
the three models tried here all gave similar $\chi^{2}/\nu$.
Therefore, the fit goodness alone cannot 
tell which model is most appropriate
in describing the spectra of X-1.
Accordingly, below we examine each model for its physical consistency,
utilizing in particular  the variation as a key probe.

The MCD+PL model reproduces the data well 
in both brighter and fainter phases,
and the inferred disk temperature
($T_{\mathrm{in}}$=1.3--1.6~keV) is similar to those
seen in high-temperature ULXs.
However, the contribution of the MCD component is only 20--50\%;
this is much smaller than is seen in high-temperature ULXs 
where the MCD component dominates their X-ray spectra.
This leads to a small inferred disk radius ($\sim 100~{\rm km}$),
and as a consequence,
the implied Schwarzschild black-hole mass 
(with $R_{\mathrm{in}}$ identified with three times the
Schwarzschild radius $R_{\mathrm{s}}$) becomes
more than an order of magnitude less than 
that derived employing the observed luminosity
and consistency with the Eddington limit.
In addition, the dominance of the PL component
at low energies (figure~\ref{fig:spectra}) would
cause difficulties in understanding the spectrum
in an analogy to the BHB's soft state,
as discussed, e.g., by \citet{Roberts2005}.
Therefore, the MCD+PL modeling fails to give
a satisfactory physical interpretation.

Although we employed the $p$-free disk model 
to examine the slim-disk interpretation, 
the derived results appear to be self-contradictory to this view.
It yields somewhat worst $\chi^{2}/\nu$ for both the fainter 
and brighter phase spectra,
together with unusually high innermost disk temperature, 
$T_{\mathrm{in}}$, exceeding 2 keV.
In addition, the $p$-free model fits imply
that the temperature profile coefficient $p$ increases
and  $T_{\mathrm{in}}$ decreases
as the source gets brighter;
these dependences are both opposite to theoretical predictions
by the slim disk model, 
and disagree with previous observations of BHBs/ULXs 
which are thought to be in the slim disk state
(e.g., \cite{Mizuno2001}; \cite{Kubota2004b}).
Therefore, X-1 is not likely to be in the slim disk state.

As shown in figure~\ref{fig:specRatio},
the X-1 spectrum below 1~keV stayed  nearly constant,
and only the hard part of the spectrum varied.
This behavior is most successfully represented by 
the  third modeling, namely MCD+cutoff-PL, making it promising.
In fact, the MCD component  in this fit is consistent 
with being constant (table~\ref{tab:time-sort}, 
figure~\ref{fig:spectra_x1}),
and the variation in the cutoff-PL component  alone 
can account for  the intensity  and spectral changes.
The cutoff-PL component represents a power-law like, but slightly convex, 
continuum shape.
Such a slightly convex hard continuum
has been recently observed from many apparently PL-like ULXs, e.g.;
IC~342 X-1 by \citet{Kubota2002};
Holmberg IX X-1 by \citet{Dewangan2006};
Holmbelg II X-1 by \citet{Goad2006};
many of the sample of 13 ULXs studied by \citet{Stobbart2006};
4 of the objects studied by \citet{Winter2005};
and NGC~1313 X-1 itself by \citet{Dewangan2005}.
Since these ULXs often  exhibit a soft spectral excess simultaneously, the
``soft excess plus power-law with high energy cutoff''
spectral shape is considered rather
common among a certain class of luminous ULXs.
In addition, the same cutoff-PL modeling was successfully employed 
by \citet{Murashima2005} to explain the variable soft X-ray
excess of the narrow-line Syfert 1 galaxy Ton~S180.
As the cutoff-PL model generally approximates 
the process of unsaturated Comptonization,
the slightly convex continua  from these high-accretion-rate  black holes
may be interpreted as a result of Comptonization of some soft photons
by hot (in this case, a few keV) thermal electrons,
which presumably form a disk corona.

Then, how can we interpret the MCD  parameters
derived from the MCD+cutoff-PL fit?
The MCD component is considered to be emitted 
by a cool disk with a temperature of  $\sim 0.2$~keV.
The derived face-value radius of 
$R_{\mathrm{in}} \sim 4000$ km (table~\ref{tab:time-sort}),
if identified with  $3R_{\mathrm{s}}$,
implies a black hole  of several hundred solar masses.
However, the luminosity attributed to the MCD component
is only a few times $10^{39}$ erg s$^{-1}$
($\sim$ 10\% of the total luminosity; table~\ref{tab:time-sort}),
which is an order of magnitude lower than the corresponding Eddington limit.
This appears to contradict to our basic assumption
that ULXs are under high accretion rates.
Furthermore, the value of $R_{\mathrm{in}}$ we obtained is 
3--4 times smaller than that measured by \citet{Miller2003}
from the same source in 2000 using XMM-Newton
(although the errors of the Suzaku measurement are quite large
due to the strong cutoff-PL component).
Similarly, \citet{Dewangan2005} analyzed 
three XMM-Newton observations of X-1,
and reported lack of the luminosity-correlated temperature changes
in the soft component which would take place if the emission is
emerging from a standard accretion disk.
These pieces of evidence, taken altogether, suggests
that the optically-thick disk, if present, has a much larger inner radius than $3 R_{\mathrm{s}}$,
or that the disk temperature is unusually low for the accretion rate.

A plausible explanation of such a  cool disk has been
obtained through the study of the Galactic BHB, XTE J1550--564,
in the very high state (\cite{Kubota2004a}; \cite{Done2005}).
In this state, a large fraction of the gravitational energy is thought to be
dissipated from a hot corona rather than from the optically-thick disk. 
Consequently, it is considered that
the accretion disk is truncated at a certain large radius
if the disk and corona are energetically independent (\cite{Kubota2004a}),
or that the disk becomes much cooler if it is strongly coupled to the corona
which takes up most of the gravitational energy release (\cite{Done2005}).
Then, the successful MCD+cutoff-PL fit to the X-1 spectra
leads to a view that the  source is in the very high state,
hosting a cool disk and a hot Comptonizing corona.
The same very-high-state interpretation,
but invoking lower accretion rates,
may also apply to the X-1 spectra obtained with  XMM-Newton  
by \citet{Miller2003}, \citet{Dewangan2005}, and \citet{Feng2006},
when the source was fainter than in the present Suzaku observation.

Since the cool disk can no longer be 
regarded as extending down to 3$R_{\rm s}$,
we cannot utilize the MCD parameters in estimating the black-hole mass.
Instead, a relatively robust estimate on the BH mass may be
obtained from the source luminosity. 
The bolometric luminosity inferred from the MCD+cutoff-PL fit 
to the Suzaku brighter-phase data of X-1
is $3.3 \times 10^{40}~{\rm erg~s^{-1}}$,
requiring a BH mass of $\sim 200~\MO$ to satisfy the Eddington limit. 
Even if we allow a super-Eddington condition by a factor of 3,
a mass of $70~\MO$ is necessary, 
which has never been observed in Galactic or Magellanic BHBs.

\subsection{Spectral State of X-2}
\label{sec:discuss_x2}
The fainter phase spectrum of X-2 is rather convex,
and is successfully reproduced by a single MCD model.
Therefore, X-2 in this phase may be regarded as a hot-MCD type ULX,
just as it was so in the two ASCA observations 
(\cite{Makishima2000}; \cite{Mizuno2001}).
When plotted on the temperature-luminosity diagram of ULXs
(figure 4 of \cite{Makishima2000}),
the present fainter-phase data point of X-2,
after the correction for the adopted distance to the source,
just falls on the line 
connecting the two ASCA data points of this source.
Since these hot-MCD type ULXs are successfully 
interpreted as residing in the slim-disk state (\S~1),
it is natural to consider, as a working hypothesis,
that  X-2 is also in the slim-disk state at least during its fainter phase.

The brighter-phase Suzaku spectrum of X-2 has a flatter shape 
as shown by figure~\ref{fig:specRatio}.
If we force a single MCD model to fit this spectrum,
we obtain $T_{\mathrm{in}} = 1.50\pm0.04$~keV 
and $R_{\mathrm{in}} = 80\pm4$~km,
even though the fit is relatively poor ($\chi^{2}/\nu=172.0/144$).
With these values,  
the brighter-phase data point falls close to the ASCA 1993 data point
on the same temperature-luminosity diagram.
Thus, the four data points of X-2,
two from ASCA and two from Suzaku,
define a common locus on the diagram,
which is characterized by
$T_{\mathrm{in}} \propto R_{\mathrm{in}}^{-1}$
(or equivalently, $L_{\mathrm{disk}} \propto T_{\mathrm{in}}^{2}$).
This relation is just what is
predicted by the slim disk model (\cite{Watarai2000}).
A more solid evidence for the slim disk is provided by the fact
that the $p$-free disk model fit to the brighter phase spectrum
gives the temperature coefficient as $p \sim 0.63$,
which is  significantly smaller than that of the standard disk ($p=0.75$),
and is consistent with what is predicted for a slim disk (\cite{Watarai2000}).
From these arguments, we conclude that X-2 was
in the slim-disk state throughout the present observation.
Although alternative interpretations 
based on the other two modelings cannot be ruled out,
the  $p$-free fit to the brighter phase data provides a most natural 
extension to the successful MCD fit  to the fainter phase spectrum.

If we assume that X-2 is shining close to the Eddington limit,
the observed luminosity requires a black-hole mass of $\sim 50~\MO$.
The value of $R_{\rm in} \sim 100$ km,
estimated using a simple MCD model, appears by
a factor of $\sim 5$ smaller than that inferred from the BH mass.
This could be due either to a high BH spin parameter,
and/or to the characteristics of slim disks,
as argued originally by \citet{Makishima2000} and \citet{Mizuno2001}, respectively.

In the XMM-Newton observation conducted in 2000 (\cite{Miller2003}),
X-2 was a factor of 2 less luminous than 
in the fainter phase observed with Suzaku.
The XMM-Newton data were reproduced by a power-law model
with $\Gamma=1.8$ plus a soft excess emission,
suggesting that the source was in the very high state.
This is reasonable, because a few Galactic BHBs show 
slightly higher luminosities (and probably much higher accretion rates) 
in the slim-disk state than 
they are in the very high state (e.g., \cite{Kubota2004b}).

In closing this subsection, let us briefly compare the two ULXs.
When time averaged,  
X-1 that is  presumably in the very high state
is about 4 times more luminous, 
than X-2 that is interpreted to be in the slim-disk state.
However, we cannot assign a higher {\em normalized} luminosity 
$\eta \equiv L_{\rm X}/L_{\rm Ed}$ 
(with $L_{\rm Ed}$ being the  Eddington limit) to X-1,
since the slim-disk state is thought to appear at a higher,
or at least comparable, value of $\eta$ 
as compared to the very high state (\cite{Kubota2004b}).
We may hence conservatively assume
that the two ULX have comparable values of $\eta$.
Then, neglecting any effect due to possible differences in their inclination,
X-1 is inferred to have (at least) 4 times higher Eddington luminosity,
and hence 4 times higher mass, than  X-2.
Then, even employing an extreme assumption 
that  X-2 has a rather ordinary stellar mass,
e.g., $\sim 10~\MO$ that requires $\eta \sim 4$,
X-1 is inferred to have $\sim 40~\MO$,
indicating that UXLs have a rather broad mass spectrum.
Of course, a more reasonable view invoking $\eta \sim 1$ would be
to assign $\sim 200~\MO$ to X-1 (\S~4.1) and $\sim 50~\MO$ to X-2.

\subsection{Oxygen Abundance}

The present Suzaku observation has for the first time provided evidence 
that the oxygen abundance in the absorption intrinsic to X-1
is significantly lower than the solar value, at 
$0.68\pm0.13$ relative to that given by \citet{Anders1982}.
This translates to
an oxygen to hydrogen number ratio of
$\mathrm{O/H} = (5.0\pm1.0) \times 10^{-4}$.
This value is consistent with that of the 
interstellar medium (ISM) in our Galaxy,
according to a recent compilation by \citet{Wilms2000}.
It is thus likely that the oxygen abundance
in the ISM is subsolar in NGC~1313 as well as in the Milky Way.
We also note that recent measurements of the solar oxygen abundance
are reconciled with the ISM oxygen abundance,
as discussed by \citet{Baumgartner2006}.

\section{Summary}

Using the Suzaku XIS, we observed the two ULXs, X-1 and X-2, in NGC~1313.
Both sources varied by about 50\% in 90 ks,
and showed intensity-correlated spectral changes.
The brighter source, X-1, exhibited the highest luminosity,
$2.5 \times 10^{40}~{\rm erg~s^{-1}}$ (0.4--10 keV),
ever recorded from this source.
The intensity and spectral variations of X-1 are
both ascribed to a strong power-law like
component with a mild high energy curvature,
while about 10\% of the flux is carried by a stable soft  component
which can be modeled by a cool ($\sim 0.2$ keV) disk emission.
These properties
suggest that the source was in the ``very high'' state,
wherein the emission is  dominated by photons
Comptonization in a hot corona,
whereas the optically-thick disk is truncated at a large radii
or cooled off as the accretion energy is taken up by the corona.
The observed properties of X-2 are consistently interpreted 
by presuming that it was in the slim disk state.
The higher luminosity of X-1 than X-2 suggests
that the black hole in X-1 is at least $\sim 4$ times
as massive as that in X-2.

\bigskip

We would like to thank N. White and Suzaku managers
for valuable comments.
We are also grateful to K. Hayashida for helpful discussions.
We also thank all the Suzaku team members
for their dedicated support of the satellite operation
and calibration.
This work was partially supported by the Grant-in-Aid for
Scientific Research of
Japan Society for the Promotion of Science (No. 18740154).

\clearpage

\begin{table*}
\caption{Time-averaged ULX spectra fitted with the
MCD+PL model}\label{tab:time-av}
\begin{center}
\begin{tabular}{cccccccc}
\hline \hline
 & $N_{\mathrm{H}}$\footnotemark[$*$] & $T_{\mathrm{in}}$ 
 & $f_{\mathrm{disk}}$\footnotemark[$\dagger$]
 & $\Gamma$ & $f_{\mathrm{pow}}$\footnotemark[$\ddagger$]
 & $L_{\mathrm{X}}$\footnotemark[$\S$] & $\chi^{2}/\nu$ \\
source & ($10^{21}~{\rm cm^{-2}}$) & (keV) 
& & & & & \\
 \hline
X-1\footnotemark[$\|$]
& $2.1\pm0.3$ & $1.60\pm0.15$ & 6.13 & $2.08^{+0.17}_{-0.21}$ & 6.27 & 25.4
& 256.9/218
\\
X-2 & $1.1^{+0.7}_{-0.2}$ & $1.29\pm0.08$ & 2.42 & $1.17(\le1.89)$ & 0.83 &
5.8 & 152.2/142 \\ \hline
\multicolumn{7}{@{}l@{}}{\hbox to 0pt{\parbox{180mm}{\footnotesize
Notes.  Errors are calculated for 90\% confidence for one
interesting parameter ($\Delta \chi^{2}=2.7$)
\par\noindent
\footnotemark[$*$] Hydrogen column density of the
photoelectric absorption 
\par\noindent
\footnotemark[$\dagger$] observed 0.4--10~keV flux of the MCD component
in units of $10^{-12}~{\rm erg~s^{-1}~cm^{-2}}$
\par\noindent
\footnotemark[$\ddagger$] observed 0.4--10~keV flux of the PL component
in units of $10^{-12}~{\rm erg~s^{-1}~cm^{-2}}$
\par\noindent
\footnotemark[$\S$] X-ray luminosity in 0.4--10~keV
in units of $10^{39}~{\rm erg~s^{-1}}$ after removing the absorption
\par\noindent
\footnotemark[$\|$] Data below 0.6~keV and
in 1.5--2.3~keV band are not used for the spectral fitting.
}\hss}}
\end{tabular}
\end{center}
\end{table*}

\begin{table*}
\caption{Fits to time-sorted spectra with errors
for 90\% confidence for one interesting parameter}\label{tab:time-sort}
\begin{center}
\begin{tabular}{lcccccccc}
\hline \hline
model &  $N_{\mathrm{H}}$ & $T_{\mathrm{in}}$ & $R_{\mathrm{in}}$
& $\Gamma$ or $p$ 
& $E_{\mathrm{c}}$ 
& $f_{\mathrm{disk}}$\footnotemark[$*$] 
& $f_{\mathrm{pow}}$\footnotemark[$\dagger$] & $\chi^{2}/\nu$ \\
       & ($10^{21}~{\rm cm^{-2}}$) & (keV) & ($\frac{1}{\cos i}$km) 
& & (keV)
& & \\ \hline
\multicolumn{9}{c}{X-1\footnotemark[$\ddagger$]} \\ \hline
fainter phase & & & & & & & \\
\hspace{0.15cm} MCD+PL & $2.4\pm0.7$ & $1.30^{+0.65}_{-0.36}$ 
 & $89^{+106}_{-54}$
 & $2.09^{+0.41}_{-0.26}$ & & 2.32 & 10.37 & 82.5/94 \\
\hspace{0.15cm} $p$-free & $2.3\pm0.2$ & $2.69^{+0.88}_{-0.38}$ 
& $21 (\le 31)$
& $0.514\pm0.014$ 
& & 12.47 & & 83.9/95 \\
\hspace{0.15cm} MCD+cutoff-PL & $3.0^{+2.1}_{-1.5}$ & $0.20(\le 0.41)$ 
& $4500 (\le 25000)$
& $1.59^{+0.33}_{-0.60}$ & $6.04^{+6.84}_{-2.27}$ & 1.92 & 11.72 & 82.1/93 \\
brighter phase & & & & & & & \\
\hspace{0.15cm} MCD+PL & $2.0\pm0.4$ & $1.63^{+0.13}_{-0.17}$ 
& $107^{+23}_{-11}$
& $2.03\pm0.24$ & & 8.31 & 8.00 & 240.4/218 \\
\hspace{0.15cm} $p$-free & $1.8\pm0.2$ & $2.09\pm0.10$ 
 & $60\pm7$
 & $0.595\pm0.012$ & & 15.58 & & 243.4/219 \\
\hspace{0.15cm} MCD+cutoff-PL & $2.3^{+0.5}_{-0.8}$ & $0.21^{+0.13}_{-0.03}$ 
& $3800^{+6700}_{-3200}$
& $0.89\pm0.20$ & $3.41^{+0.57}_{-0.40}$ & 1.48 & 15.13 & 237.9/217 \\
\hline
\multicolumn{9}{c}{X-2} \\ \hline
fainter phase & & & & & & & \\
\hspace{0.15cm} MCD & $0.9\pm0.3$ & $1.24\pm0.05$ 
 & $96\pm9$
 & & & 2.28 & & 71.2/76 \\
brighter phase & & & & & & & \\
\hspace{0.15cm} MCD+PL & $1.6^{+0.7}_{-0.4}$ & $1.39^{+0.24}_{-0.14}$ 
& $80^{+30}_{-20}$
& $1.62^{+0.58}_{-1.09}$ & & 2.59 & 1.61 & 148.9/142 \\
\hspace{0.15cm} $p$-free & $1.7\pm0.4$ & $1.86\pm0.15$ 
 & $43^{+12}_{-9}$
 & $0.627^{+0.036}_{-0.026}$ & & 4.20 & & 150.8/143 \\
\hspace{0.15cm} MCD+cutoff-PL & $2.1\pm0.9$ & $0.24(\le0.57)$ 
& $1300 (\le 4600)$
& $0.59^{+0.27}_{-0.41}$ & $2.61^{+0.58}_{-0.47}$ & 0.36 & 4.01 & 147.6/141 \\
\hline
\multicolumn{9}{@{}l@{}}{\hbox to 0pt{\parbox{180mm}{\footnotesize
\par\noindent
\footnotemark[$*$] observed 0.4--10~keV flux 
of the MCD or the p-free disk component
in unit of $10^{-12}~{\rm erg~s^{-1}~cm^{-2}}$
\par\noindent
\footnotemark[$\dagger$] observed 0.4--10~keV flux of the cutoff PL component
in unit of $10^{-12}~{\rm erg~s^{-1}~cm^{-2}}$
\par\noindent
\footnotemark[$\ddagger$] Data below 0.6~keV and
within the range 1.5--2.3~keV are not used for the spectral fitting.
}\hss}}
\end{tabular}
\end{center}
\end{table*}

\clearpage

\end{document}